\documentclass[letterpaper, 10 pt, conference]{ieeeconf}  
\usepackage{color}
\usepackage{graphicx}
\usepackage{multirow}
\usepackage{amssymb}
\usepackage{array}
\newcommand{\tabincell}[2]{\begin{tabular}{@{}#1@{}}#2\end{tabular}}

\IEEEoverridecommandlockouts                              

\title{\LARGE \bf Low-latency auditory spatial attention detection based on spectro-spatial features from EEG}

\author{Siqi Cai$^{1,*}$, Pengcheng Sun$^{1,*}$ Tanja Schultz$^{2}$, and Haizhou Li$^{1,3}$
\thanks{This work has been submitted to the IEEE for possible publication. Copyright may be transferred without notice, after which this version may no longer be accessible.}
\thanks{$^{1}$Siqi Cai, Pengcheng Sun and Haizhou Li are with the Department of Electrical and Computer Engineering, National University of Singapore, Singapore. Siqi Cai is the corresponding author.
 {\tt\small elesiqi@nus.edu.sg}, {\tt\small pengcheng.sun@u.nus.edu}, and {\tt\small haizhou.li@nus.edu.sg}}%
\thanks{$^{2}$Tanja Schultz is with Cognitive Systems Lab, University of Bremen, Germany.
{\tt\small tanja.schultz@uni-bremen.de}}%
\thanks{$^{3}$Haizhou Li is also with Machine Listening Lab, University of Bremen, Germany. }%
 \thanks{${*}$Equal contribution}%
}

\begin{document}
\maketitle
\thispagestyle{empty}
\pagestyle{empty}
\makeatletter

\begin{abstract}

Detecting auditory attention based on brain signals enables many everyday applications, and serves as part of the solution to the cocktail party effect in speech processing. Several studies leverage the correlation between brain signals and auditory stimuli to detect the auditory attention of listeners. Recently, studies show that the alpha band (8-13 Hz) EEG signals enable the localization of auditory stimuli. We believe that it is possible to detect auditory spatial attention without the need of auditory stimuli as references. In this work, we use alpha power signals for automatic auditory spatial attention detection. To the best of our knowledge, this is the first attempt to detect spatial attention based on alpha power neural signals. We propose a spectro-spatial feature extraction technique to detect the auditory spatial attention (left/right) based on the topographic specificity of alpha power. Experiments show that the proposed neural approach achieves 81.7\% and 94.6\% accuracy for 1-second and 10-second decision windows, respectively. Our comparative results show that this neural approach outperforms other competitive models by a large margin in all test cases. 

\end{abstract}

\section{INTRODUCTION}
Humans have the ability to pay attention to particular sound sources and voices, even in multi-talker scenarios~\cite{cherry1953some}
- the so called cocktail party effect. Previous studies have revealed the role of specific neural processes involved, and provided neural evidence for auditory attention modulation~\cite{kerlin2010attentional,mesgarani2012selective,o2015attentional}. With the latest advancements in neuroscience, we are inspired to develop computational models that detect the auditory attention as part of the brain activities.

Recent findings show that auditory attention in cocktail party scenarios can be decoded from the recordings of brain activity, such as electrocorticography (ECoG)~\cite{mesgarani2012selective}, magnetoencephalography (MEG)~\cite{ding2012neural} and electroencephalography (EEG)~\cite{o2015attentional,das2016effect,de2017machine,ciccarelli2019comparison,cai2020low,jaeger2020decoding}. Among them, EEG provides a non-invasive and low-cost
means of investigating cortical activity with high temporal resolution, which makes it particularly suitable for brain-computer interface (BCI) applications~\cite{hwang2013eeg}. Therefore, we are interested in the decoding of the auditory attention from EEG signals in this paper.

Most of the studies on auditory attention detection are focused on detecting the envelope of the speech produced by the attended speaker, that is referred to as speech envelope reconstruction technique. Such technique requires the auditory stimulus, i.e. the clean speech signal recorded in a noise-free environment, to be available~\cite{o2015attentional,das2016effect,de2017machine,miran2018real}. Unfortunately, in the real-world applications, such as hearing prostheses or speaker localisation, it is unrealistic to obtain such clean speech signals. Inspired by the findings that alpha power is highly associated with spatial attention~\cite{wostmann2016spatiotemporal,bednar2020cocktail,deng2020topographic}, we hypothesize that we can detect the auditory spatial attention based on brain activities alone, without the need of clean speech envelopes. 

Meanwhile, it was shown that the linear EEG decoder requires a very long decision window, with duration of 10 seconds or more, for a reliable decision on auditory spatial attention~\cite{miran2018real}. A response delay of 10 seconds is out of question for applications such as hearing aids. Thus, non-linear decoders with shorter decision windows are of high interest. The latest deep learning techniques provide new ways to understand the complex and highly non-linear nature of auditory processes in human brain. Non-linear decoders~\cite{de2017machine,ciccarelli2019comparison,deckers2018eeg,vandecappelle2020eeg} have shown superior performance to linear decoders in several low-latency settings. In this paper, we further the study of a non-linear decoder for low latency auditory spatial attention detection (ASAD).

\begin{figure*}[htb]
  \centering
  \setlength{\abovecaptionskip}{0cm}
  \includegraphics[width=0.95\linewidth]{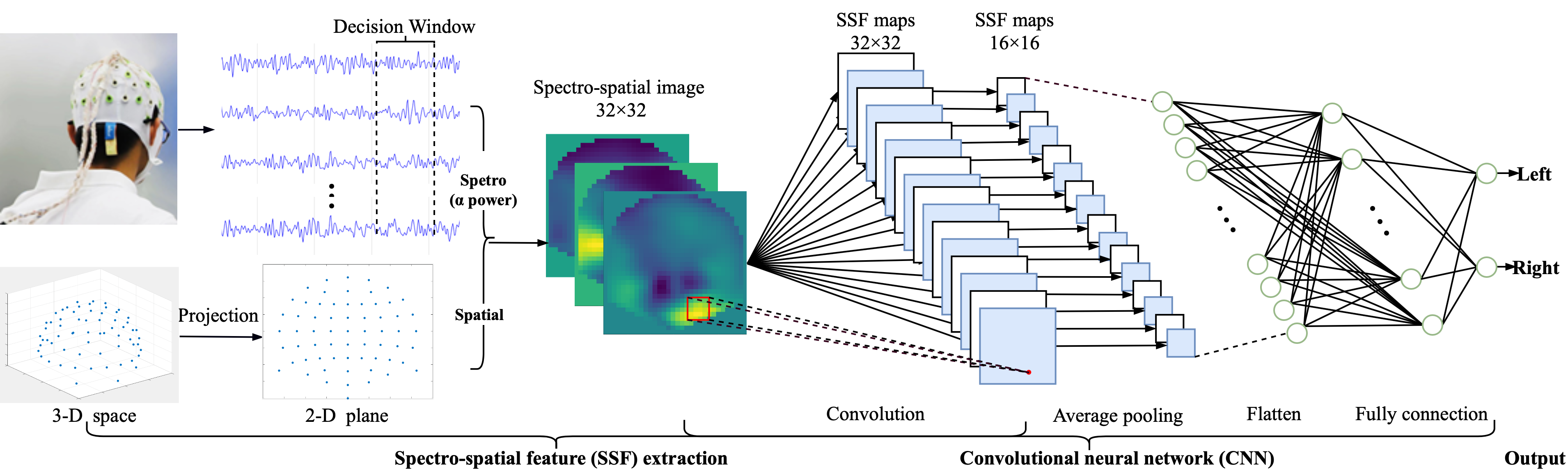}
  \caption{The proposed convolutional neural network (CNN) with spectro-spatial feature (SSF) for auditory spatial attention detection, that is referred to as SSF-CNN model. The SSF-CNN network is trained to output two values, i.e., 0 and 1, to indicate the spatial location of the attended speaker.}
  \label{CNN}
  \vspace{-0.5cm} 
\end{figure*}

The contributions of this paper come in three parts: (1) the design, implementation, extraction, as well as combination of spectral plus spatial features from the EEG alpha band to form spectro-spatial feature (SSF), (2) the application of convolutional neural network (CNN) based classification of auditory spatial attention, and (3) the combination of these two components to form the SSF-CNN system for ASAD, as illustrated in Fig.~\ref{CNN}. The final SSF-CNN system is experimentally evaluated and outperforms other competitive models in both accuracy and latency.

\section{Auditory Spatial Attention Detection}

\subsection{Spectro-spatial Feature (SSF)}
The topological distribution of oscillatory cortical activity in the alpha frequency band is closely related to the location of spatial focus of attention \cite{deng2020topographic}, that prompts us to study a novel spectro-spatial feature extraction technique. We study the feature extraction in two stages as shown in Fig.~\ref{CNN}.

First, a fast Fourier transform (FFT) is performed on the continuous time series of each electrode to obtain the power spectrum of the EEG signal. The average of squared absolute value in the frequency band is then taken as the individual measurement value of each electrode.

Second, we propose to convert these measurements of different decision windows into a sequence of 2-D images, so as to take full advantage of the spatial features of EEG signals. In this stage, EEG electrodes are projected from the 3-D space onto a 2-D plane according to the coordinate information using Azimuth Equidistant Projection~\cite{snyder1987map}. In practice, we project all points onto a plane tangent to the earth, and divide all the latitude and longitude lines into equal parts, which ensures that all points are accurately spaced and oriented from the center. Considering a human head approximately as a sphere, we select the top point of the head as the tangent point,
so as to obtain the projection of the electrode on the 2-D image. The Clough-Tocher interpolant~\cite{amidror2002scattered}, which is based on cubic polynomial, is used to estimate the value of each grid over a 32$\times$32 mesh, which represents the spatial distribution of EEG signals. 

In this way, a topographical activity map of EEG signals can be generated, that represents the alpha frequency band within a time window. As the map takes both spectral and spatial information of EEG data into account, it is referred to as the SSF map.
We then use the sequence of SSF maps derived from consecutive time windows to reflect the temporal evolution of brain activities, which serve as the input to the subsequent convolutional neural network.

Overall, the proposed SSF extraction facilitates the learning of the topographic specificity of alpha power from EEG signals~\cite{deng2020topographic}. We have good reason to expect that SSF is more expressive than the original EEG signals in attention detection.
Moreover, it eliminates the need for handcrafting any features.

\subsection{Attention Detection with Convolutional Neural Network}

Convolutional neural network (CNN) is a kind of feedforward neural network, which makes use of convolution and pooling techniques for representation learning and classification decision. The fact that CNN is effective in image recognition~\cite{de2017detecting} leads us to believe that it would also learn and classify well the sequence of topographical EEG maps.

As shown in Fig. \ref{CNN}, the CNN architecture starts with a convolution layer, which uses a kernel size of 3$\times$3 and a stride of 1 with padding. The convolution layer has a rectifying linear unit (ReLU) activation function, and is followed by an average pooling layer with a 2$\times$2 kernel and 2 pixels stride, and two fully-connected layers with 512 and 32 neurons, respectively. The batch normalization is applied to every convolution layer to reduce the effect of the distribution of internal neurons. To avoid overfitting, a dropout layer~\cite{hinton2012improving} is applied after the pooling layer and the first fully connection layer, respectively. Finally, a softmax output layer is added for binary decision. 

Cross-entropy loss is selected as the cost function, using the root mean square propagation algorithm (RMSProp)~\cite{kurbiel2017training}. Both the learning rate and decay are set to 1$\times$10$^{-3}$.

\section{Experiments and Results}

\subsection{Experimental Setup} 
We conducted the auditory attention detection experiments on the dataset recorded at KU Leuven~\cite{das2019auditory}, denoted as KUL Dataset. Briefly, 64-channel EEG data was collected from eight male and eight female normal-hearing subjects. A subject was instructed to pay attention to one of two competing speakers. The EEG data was recorded with a BioSemi ActiveTwo device at a sampling rate of 8,192 Hz and an electrode positioning that follows the international 10-20 system. Four Dutch short stories, narrated by different male speakers, were used as the stimuli. The auditory stimuli were low-pass filtered with a cut-off frequency of 4 kHz and presented at 60 dB through a pair of in-ear earphones (Etymotic ER3).

The experiment for each subject was split into eight trials of 6 minutes duration. The auditory stimuli were either presented dichotically (one speaker per ear) or with two speakers coming from 90 degrees to the left and 90 degrees to the right of the subject, respectively. The latter stimuli was simulated based on a head-related transfer function (HRTF) filtering. Throughout the experiments, the order of presentation of both conditions was randomized over the different subjects. In total, 8 × 6 min = 48 min of EEG data was collected for each subject, accumulating to 12.8 hours of EEG data for all 16 subjects.

\subsection{Data Preparation}

The EEG data of each channel were re-referenced to the average response of the mastoid electrodes, then bandpass-filtered between 8 and 13 Hz, and subsequently down-sampled to 70 Hz. The frequency range was chosen based on non-linear auditory attention detection studies~\cite{de2017machine,deckers2018eeg,vandecappelle2020eeg}. Finally, EEG data channels were normalized to ensure zero mean and unit variance for each trial.

The data set was randomly split into a training (80\%), a validation (10\%), and a test set (10\%) while preserving the distribution between left/right attention in the three partitions by subject. For each partition, the data segments were generated with a sliding window (referred to as \textit{decision window}) with an overlap of 50\%. Thus, for the 0.1-second decision window, the test set resulted in 5,760 decision windows per subject, totaling to 92,160 decision windows.

\subsection{Experiments with EEG of 64 Channels}

\begin{figure*}[t]
      \centering
      \includegraphics[width=0.75\linewidth]{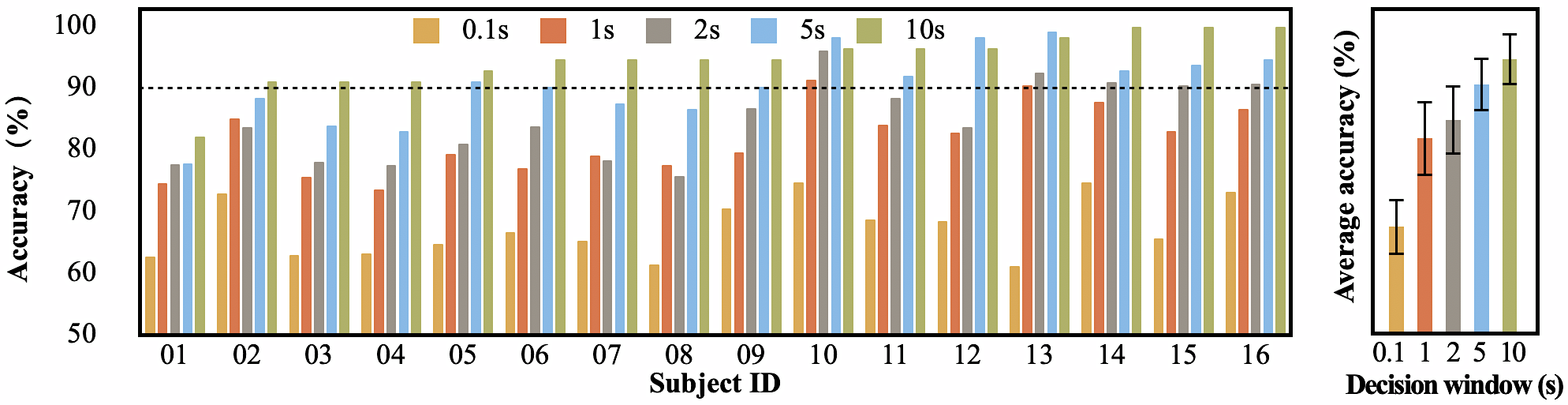}
      \caption{Auditory spatial attention detection accuracy of the proposed SSF-CNN model with 64-channel EEG over different decision windows and for all subjects. The subjects are ranked according to the accuracy for the 10-second decision window. The horizontal dotted line shows a reference point of high accuracy at 90\%.}
      \label{result}
      \vspace{-0.4cm} 
   \end{figure*}

The SSF-CNN model is an end-to-end network, which decides between left and right attention for each EEG data segment. As the test set is balanced between left-right attention, the chance-level is 50\%. To avoid initialization bias, the accuracy is averaged over 10 runs with random initialization. We report the overall average detection accuracy and the average accuracy per subject for five decision window sizes ranging from 0.1 to 10 seconds. The results are presented in Fig.~\ref{result} and show an accuracy of 67.2\% (SD: 4.57) for 0.1-second, 81.7\% (SD: 5.37) for 1-second, 84.7\% (SD: 6.13) for 2-second, 90.5\% (SD: 5.71) for 5-second, and 94.6\% (SD: 4.37) for the 10-second decision window. Overall, it is apparent that longer decision windows lead to higher detection accuracy. While there are exceptions where longer decision windows do not help, the accuracy trend over window size corroborates with findings in other studies~\cite{de2017machine,ciccarelli2019comparison,cai2020low,jaeger2020decoding}. It is worth noting that SSF-CNN shows a more consistent accuracy trend over window size than the CNN~\cite{vandecappelle2020eeg} baseline, with a fewer number of exceptions.

To assess the performance of our proposed SSF-CNN system we used two benchmarks, a linear decoder baseline~\cite{o2015attentional} and a non-linear CNN model~\cite{vandecappelle2020eeg}. For the former, we re-implemented the stimulus (speech envelope) reconstruction model~\cite{o2015attentional}, in which the EEG signals approximate the envelope of the attended speech. The reconstructed stimulus is then compared with the original by calculating the correlation between them. Strong correlation indicates the presence of attention. For the latter, we refer to the results reported by Vandecappelle et al.~\cite{vandecappelle2020eeg}, who performed auditory attention detection experiment on the same KUL Dataset. The authors applied a non-linear CNN model with a $\left[64,T\right]$ matrix as input, that represents 64 EEG channels and $T$ samples in a decision window. While our SSF-CNN model leverages our proposed spectro-spatial features as input to the CNN.  

Table I shows the auditory spatial attention detection accuracy of the proposed SSF-CNN model in comparison to the two benchmark models. The SSF-CNN models consistently outperforms both the linear~\cite{o2015attentional} and the CNN~\cite{vandecappelle2020eeg} model. The differences become more prominent with shorter window length. In particular, the linear decoder accuracy drops significantly when operating on short decision windows, i.e., 58.1\% with 1-second decision window, while the SSF-CNN model (81.7\%) maintains the performance level reasonably well. The variation of decoding accuracy with window length is consistent with the literature~\cite{de2017machine,cai2020low,deckers2018eeg,vandecappelle2020eeg}. 

We carried out a significance test to confirm that the SSF-CNN model improves significantly (paired \textit{t}-test, \textit{p} = 0.039) over its CNN counterpart~\cite{vandecappelle2020eeg}. Since the major difference between the models lies in the EEG feature representation, we believe that the performance improvements are in fact a result of the topographic specificity of alpha power signals, which acts as a spatially selective filter of attention in cocktail party scenarios~\cite{wostmann2016spatiotemporal,bednar2020cocktail,deng2020topographic}.

Furthermore, while the 1-second decision window is close to the time lag required by humans to switch attention~\cite{miran2018real}, we wanted to push the limits with regards to the decision window lengths. Thus, we tested SFF-CNN and the CNN model with window length that is shorter by an order of magnitude, i.e., 100 millisecond (ms). It is encouraging to see that the SSF-CNN model not only outperforms the CNN model at same window length, but also the linear model with 1-second, and 2-second window lengths. 

To the best of our knowledge, the SSF-CNN model achieves the best accuracy on KUL dataset with all decision windows ranging from 0.1 to 10 seconds.
Since it eliminates the need for a reference auditory stimulus, the proposed SSF-CNN model represents a very appropriate solution for neuro-steered hearing prostheses and other everyday applications, and remains viable even for low-latency requirements. 

\begin{table}[]
\begin{center}
\caption{Attention detection accuracy (\%) on KUL Dataset of 64 and 32 channel EEG (\#EEG) for five decision window sizes.}
\begin{tabular}{|l|c|c|p{0.4cm}<{\centering}|p{0.4cm}<{\centering}|p{0.4cm}<{\centering}|p{0.4cm}<{\centering}|p{0.4cm}<{\centering}|}
\hline
\multirow{2}{*}{Model} & \multirow{2}{*}{\#EEG}    & \multirow{2}{*}{\tabincell{c}{Auditory\\ stimulus}}      & \multicolumn{5}{c|}{Decision window (second)} \\  \cline{4-8}
                            & &  &0.1 & 1         & 2        & 5     & 10       \\ \hline
Linear \cite{o2015attentional} & 64 & with    &- & 58.1          & 61.3        & 67.5    & 75.8    \\ \hline
CNN \cite{vandecappelle2020eeg} & 64 & without  &65.9 & 80.8         & 82.1         & 83.6     & 85.6 \\ \hline
\textbf{\tabincell{c}{SSF-CNN}}   & 64 & without   &\textbf{67.2}     & \textbf{81.7}        & \textbf{84.7}       & \textbf{90.5}            & \textbf{94.6}       \\ \hline
\textbf{\tabincell{c}{SSF-CNN }}   & 32 & without   &-   &  \textbf{76.1}        & \textbf{80.1}      & \textbf{86.2}          & \textbf{89.4}      \\ \hline
\end{tabular}
\end{center}
\vspace{-0.7cm} 
\end{table}

\subsection{Experiments with EEG of 32 Channels}
Results reported so far, relied on 64-channel EEG data. Since a lower number of EEG electrodes has multiple advantages, we reduced the number of electrodes from 64 to 32 channels, following the international 10/20 system~\cite{homan1987cerebral}. 

In Table I, we compare the detection accuracy of the SSF-CNN model between 32-channels and 64-channel signals. The detection accuracy for the 32-channel version is 76.1\% (SD: 6.84), 80.1\% (SD: 7.56), 86.2\% (SD: 6.05), and 89.4\% (SD: 8.09) for 1, 2, 5, and 10-second decision windows, respectively. While the accuracy of 32-channel data is generally lower than that of 64-channel data, the mean accuracy remains around 80\% with a 2-second decision window. In addition, the 32-channel SSF-CNN model outperforms the linear model with 64-channel EEG over all decision windows lengths. From Table II, we also observe that the 32-channel SSF-CNN model compares favorably~\cite{vandecappelle2020eeg} for longer window sizes. In sum, the proposed SSF-CNN method detects the auditory spatial attention accurately even with a reduced set of EEG channels, which is an important feature for real-world application.

\section{CONCLUSIONS}
In this paper we proposed a novel spectro-spatial feature representation for EEG that serves as input into a CNN model to perform auditory spatial attention detection. The resulting SSF-CNN system consistently and significantly outperforms two benchmark models, a conventional linear model and a state-of-the-art CNN model, over various window lengths. Furthermore, the SSF-CNN achieves encouraging results even with extremely short decision window length and a reduced number of EEG channels. Most importantly, the proposed feature representation does not require any clean reference signal. The combination of these outcomes make the SSF-CNN a highly competitive candidate for real-life applications such as neuro-steered hearing aids. 

\addtolength{\textheight}{-12cm}   




\section*{ACKNOWLEDGMENT}

This research work is supported by Programmatic Grant No. A18A2b0046 and A1687b0033 from the Singapore Government’s Research, Innovation and Enterprise 2020 plan (Advanced Manufacturing and Engineering domain). 

The work by Haizhou Li and Tanja Schultz is also funded by the Deutsche Forschungsgemeinschaft (DFG, German Research Foundation) under Germany's Excellence Strategy (University Allowance, EXC 2077, University of Bremen, Germany). 


\bibliographystyle{IEEEtran}
\bibliography{mybib}

\end{document}